\documentclass[12pt,preprint]{aastex}
\newcommand{\kms}{${\rm km~s}^{-1}$}

\begin{document}

\title{Template RR Lyrae ${\rm H\alpha}$, ${\rm H\beta}$, and ${\rm H\gamma}$
velocity curves}

\author{Branimir Sesar}
\affil{Division of Physics, Mathematics and Astronomy, California Institute of
       Technology, Pasadena, CA 91125}
\email{bsesar@astro.caltech.edu}

\begin{abstract}
We present template radial velocity curves of $ab$-type RR Lyrae stars
constructed from high-precision measurements of ${\rm H\alpha}$, ${\rm H\beta}$,
and ${\rm H\gamma}$ lines. Amplitude correlations between the Balmer line
velocity curves, Johnson $V$-band, and SDSS $g$- and $r$-band light curves are
also derived. Compared to previous methods, these templates and derived
correlations reduce the uncertainty in measured systemic (center-of-mass)
velocities of RR Lyrae stars by up to 15 {\kms}, and will be of particular
interest to wide-area spectroscopic surveys such as the Sloan Digital Sky Survey
(SDSS) and LAMOST Experiment for Galactic Understanding and Exploration (LEGUE).
\end{abstract}

\keywords{stars: horizontal-branch -- stars: Population II -- stars: variables: RR Lyrae -- techniques: radial velocities}

\section{Introduction\label{introduction}}

RR Lyrae stars are old ($\gtrsim9$ Gyr), low-mass ($\sim0.7M_\sun$), pulsating
stars ($V$-band amplitudes of $A_V\sim1$ mag and periods of $\sim 0.6$ days)
that reside in the instability strip of the horizontal branch \citep{smi95}.
They are valuable tracers of old stellar populations and have recently been used
to map the Galactic halo structure (i.e., its stellar number density profile)
and substructure (e.g., tidal streams and halo overdensities; \citealt{viv01},
\citealt{kel08}, \citealt{mic08}, \citealt{wat09}, \citealt{ses10a}).

While the spatial distribution of RR Lyrae stars is a powerful indicator of
candidate halo substructures (e.g., Fig.~11 of \citealt{ses10a}), it does not
always reveal the full nature of halo substructures. For example, the Pisces
Overdensity, a spatial group of about a dozen RR Lyrae stars at $\sim80$ kpc
from the Sun (clump ``J'' in \citealt{ses07}; \citealt{wat09}), was initially
suspected to be a part of a tidal stream \citep{ses10a}. However, spectroscopic
observations by \citet{kol09} and \citet{ses10b} have revealed that the
overdensity actually consists of {\em two} velocity groups; one moving towards
and the other one moving away from the Galactic center. This result suggests
that the Pisces Overdensity may be a pile-up of tidal debris near the orbital
turn-around point of a now-disrupted progenitor (which was, most likely, a dwarf
spheroidal galaxy; \citealt{ses10b}).

The \citet{kol09} and \citet{ses10b} findings underline the importance of
obtaining kinematic data on candidate halo substructures; the nature and
interpretation of a halo substructure can quickly change once kinematic data
become available. Yet, relative to photometric data, which is abundant in this
era of wide-area and multi-epoch surveys, spectroscopic observations are more
scarce and may be considered as more ``expensive'' in terms of telescope time.
Obtaining the kinematic information for variable stars, such as RR Lyrae stars, 
is even more complicated as one needs to subtract the velocity due to pulsations
to get the center-of-mass (hereafter, systemic) velocity.

There are two approaches to this problem of determining the systemic velocity of
RR Lyrae stars. In the first approach, one tries to observe RR Lyrae stars at a
particular point during a pulsation period when the observed radial velocity is
presumed to be equal to the star's systemic velocity (usually at
${\rm phase\sim0.5\pm0.1}$, see Section~\ref{systemic_velocity}). Unfortunately,
sometimes it is very difficult to schedule observations to target this point
(e.g., when executing observations in queue or service mode). In addition, RR
Lyrae stars are dimmer near phase 0.5 ($\sim0.3-0.5$ mag compared to maximum
light at phase of 0), and thus require longer exposures to reach the same
signal-to-noise ratio. The danger of longer exposures is blurring of spectral
lines due to pulsations, which in turn decreases the precision of radial
velocity measurements.

In the second approach, one assumes a certain model (a template) that describes
the radial velocity due to pulsations as a function of pulsation phase. The
amplitude, $A_{rv}$, of the radial velocity curve template (hereafter, template)
is then scaled using a correlation between the amplitudes of velocity curves and
light curves. The velocity at any phase can then be easily read-out once the
scaled template is matched to one or several radial velocity observations taken
at known phases. An example of such a template and an equation relating velocity
and light curve amplitudes is provided by \citet{liu91} (see his Equation 1).
The advantage of this approach is that the systemic velocity can be estimated
using radial velocities obtained in phases different from phase of 0.5.

Unfortunately, Liu's template and his velocity vs.~light curve amplitude
relation are not really suitable for estimating systemic velocities of distant
(i.e., faint) RR Lyrae stars because they were derived from measurements of
{\em metallic} lines and should only be used with measurements of such lines.
This is a problem since measuring radial velocities using metallic lines
requires high-resolution spectroscopy ($R\gtrsim10000$), and obtaining quality,
high-resolution spectra of faint ($V>18$) stars with exposures that are a small 
fraction of the pulsation period ($\lesssim10\%$ to avoid spectral blurring) is
virtually impossible with today's facilities.

Instead of using weak metallic lines, a more practical choice would be to use
strong Balmer lines (${\rm H\alpha}$ to ${\rm H\delta}$), which can be observed
with more widely available low-resolution ($R\sim1500$) spectrographs. However,
Liu's template and his velocity vs.~light curve amplitude relation cannot be
used with measurements of Balmer lines because the Balmer line velocity curves
have a different shape {\em and} amplitude than the velocity curve of metallic
lines \citep{oke62}. A method adopted by many studies (e.g.,
\citealt{hb85, lay94, vzg05, pri09, ses10b, ses12}) is to use the radial
velocity curve of RR Lyrae star X Ari measured from the ${\rm H\gamma}$ line by
\citet{oke66}. This is not an ideal solution because the velocity curve
amplitude of X Ari ($A_{rv}\sim85$ {\kms}, see Figure~1 of \citealt{oke66}) will
likely not be similar to the velocity curve amplitude of some other star with a
different $V$-band light curve amplitude (${\rm A_V}$). The consequence of using
an inadequate model for radial velocities is a more uncertain estimate of the
systemic velocity.

The ability to precisely measure systemic velocities of RR Lyrae stars is very
important as it may help with various Galactic studies, such as identifying
halo substructures using kinematics (e.g., the Pisces Overdensity groups),
discerning the nature of progenitors of halo substructures (e.g., globular
cluster vs.~a dwarf spheroidal galaxy), and constraining orbits of halo
substructures (e.g., the Virgo Overdensity; \citealt{cd09, car12}). In addition,
obtaining precise systemic velocities of RR Lyrae stars using one or two
measurements would be beneficial to spectroscopic surveys such as the LAMOST
Experiment for Galactic Understanding and Exploration (LEGUE; \citealt{den12}),
which may observe several thousands of RR Lyrae stars within 30 kpc of the Sun,
and which plan to study the kinematics and structure of the Galactic halo.

In this paper, we use high-precision measurements of ${\rm H\alpha}$,
${\rm H\beta}$, and ${\rm H\gamma}$ lines of six field $ab$-type RR Lyrae stars
to first construct template radial velocity curves (Section~\ref{templates}). We
then derive correlations between amplitudes of Balmer line velocity curves and
light curves observed in Johnson $V$-band (Section~\ref{correlations}) or SDSS
$g$- and $r$-bands (Section~\ref{sdss_correlations}). The systemic velocity of
RR Lyrae stars and its uncertainty are defined in
Section~\ref{systemic_velocity}, and the results, along with implications for
surveys such as LINEAR, PTF, LEGUE, and LSST, are discussed in
Section~\ref{discussion}.

\section{Data set}

The sample of stars used in the construction of velocity templates consists of 6
field RR Lyrae type $ab$ stars (see Table~\ref{ASAS_RR}) that were observed
between 2006 and 2009 by George W.~Preston with the echelle spectrograph of the
du Pont 2.5 m telescope at the Las Campanas Observatory. The stars have $V$-band
light curve amplitudes (${\rm A_V}$) ranging from 0.64 mag (Z Mic) to 1.13 mag
(RV Oct), as measured by the All Sky Automated Survey (ASAS; \citealt{poj97}).
According to \citet{sf07}, three of the stars in this sample exhibit the Blazhko
effect in their light curves \citep{bla07,smi95}.

The observations, data reduction, and measurement of radial velocities are
described in detail in \citet{fps11}. The data, kindly provided to us by George 
W.~Preston (private communication), consist of radial velocity measurements
obtained using metallic and Balmer lines (${\rm H\alpha}$, ${\rm H\beta}$, and
${\rm H\gamma}$). The formal random error is $\sim0.5$ {\kms} for radial
velocities obtained using metallic lines, and $\sim3$ {\kms} for radial
velocities obtained using Balmer lines. In addition, systematic errors of
$\sim1$ to 2 {\kms} may also be present. The phase of each observation was
calculated using periods and epochs of maximum light listed in Table 5 of
\citet{fps11}, with the time of maximum light corresponding to phase of 0. The
number of observations per star ranges from about 100 (DT Hya) to about 240 (XZ 
Aps).

\section{Construction of velocity templates\label{templates}}

We start by normalizing radial velocities to the range ${\rm [0,1]}$. The
normalization is done on a per star basis using
\begin{equation}
rv_{norm} = \frac{rv - rv_{min}}{rv_{max} - rv_{min}},
\end{equation}
where $rv_{min}$ and $rv_{max}$ are, respectively, the lowest and highest radial
velocity determined after discarding the top and bottom 1\% of radial
velocities of that star. The normalized velocities from all stars are then
combined and a cubic B-spline is interpolated through the data. The normalized, 
phased, and combined radial velocities are shown in Figure~\ref{fig1} (top plots
in each panel). The velocities and the interpolated B-spline (the template) are 
offset so the velocity at phase 0.5 (the phase at which the radial velocity is
presumed to be equal to the systemic velocity) is equal to 0. The templates are 
provided in the supplementary data available in the electronic edition of the
journal and at \url{http://www.astro.caltech.edu/~bsesar/RVtemplates.tar.gz}.

Note that the templates of Balmer line velocity curves do not cover the full
range of phases, but end at phase of 0.95. Due to a rapid change in velocities
of Balmer lines between phases of 0.95 and 1.0, we were not able to properly
model the entire velocity curve with a single cubic spline. We do not consider
this to be a major problem from an observational point of view because the
velocities measured from Balmer lines between phases of 0.9 and 1.0 are in
general more uncertain, mainly due to increased spectral blurring (i.e., due to
a rapid change in velocity as a function of time), and are not very useful.

The interpolated cubic B-spline defines the average shape of the radial velocity
curve. To estimate the uncertainty in this average shape, we calculate the
root-mean-square (hereafter, rms) scatter around the template and plot its
dependence on phase in bottom plots in each panel of Figure~\ref{fig1}. The
shape of the ${\rm H\alpha}$ template is most uncertain, with an rms scatter of
$\sim10\%$ of ${\rm A_{rv}}$ at phase of about 0.55 ($\sim10$ {\kms} for a RR
Lyrae star with ${\rm A_V=1.0}$ mag). The shape of the metallic lines template
curve is the least uncertain ($\sim3\%$ of ${\rm A_{rv}}$), followed by
${\rm H\gamma}$ and ${\rm H\beta}$ template curves ($\sim3\%$ to $4\%$ of
${\rm A_{rv}}$ or $\sim4$ {\kms} for a RR Lyrae star with ${\rm A_V=1.0}$ mag).

\section{${\rm A_{rv}}$ vs.~${\rm A_V}$ correlations\label{correlations}}

Having defined average velocity curves (templates), we now fit templates to
observed radial velocities of each star to determine the amplitudes of metallic
and Balmer lines velocity curves (${\rm A_{rv}}$). The best-fit amplitudes are
listed in Table~\ref{ASAS_RR}, and are plotted versus $V$-band amplitudes in
Figure~\ref{fig2}.

As \citet{liu91} (hereafter, L91) first noted, there is a tight correlation
between the amplitudes of metallic lines velocity curves and $V$-band light
curves. We also see a similar correlation, but for velocity curves measured from
Balmer lines. Following L91, we do an unweighted, linear least-squares fit to
velocity amplitudes as a function of the $V$-band light curve amplitudes and
find:

\begin{eqnarray}
A_{rv}^{met} = 25.6(\pm2.5)A_V + 35.0(\pm2.3),\, \sigma_{fit}=2.4\, km\, s^{-1}\label{metallic_lines}\\
A_{rv}^{H\alpha} = 35.6(\pm2.5)A_V + 78.2(\pm2.4),\, \sigma_{fit}=3.4\, km\, s^{-1}\label{Halpha} \\
A_{rv}^{H\beta} = 42.1(\pm2.5)A_V + 51.1(\pm2.4),\, \sigma_{fit}=3.0\, km\, s^{-1}\label{Hbeta} \\
A_{rv}^{H\gamma} = 46.1(\pm2.5)A_V + 38.5(\pm2.4),\, \sigma_{fit}=2.8\, km\, s^{-1}\label{Hgamma},
\end{eqnarray}
where ${\rm \sigma_{fit}}$ is the rms scatter of the best fit.

In the top left panel of Figure~\ref{fig2}, we compare our best-fit for metallic
lines (Equation~\ref{metallic_lines}) to
\begin{equation}
A_{rv}^{L91} = \frac{40.5A_V+42.7}{1.37}\label{Liu_RV},
\end{equation}
which is the best-fit for metallic lines obtained by L91 (his Equation 1) scaled
by the so-called ``projection factor'' $p = 1.37$. Equation 1 of L91 needs to be
scaled because L91 uses {\rm pulsation} velocities, which are related to
observed radial velocities as $v_{obs}=v_{puls}/p$ (see Liu's Equation 2). The
projection factor was estimated as the average ratio of velocities calculated
using Liu's Equation 1 and metallic ${\rm A_{rv}}$ values listed in
Table~\ref{ASAS_RR}. The estimated projection factor $p=1.37$ is in the range of
values listed by L91 and \citet{kov02} (1.31 to 1.37). In conclusion, the scaled
L91 fit and our fit are remarkably similar (within $\sim2.5$ {\kms} of rms
scatter), even though Liu's sample has more stars and spans a much greater range
of $V$-band amplitudes than our sample (Liu's sample extends up to
${\rm A_V\sim1.3}$ mag).

Figure~\ref{fig2} shows that for a fixed $V$-band amplitude, each of the Balmer
lines studied here has a different velocity curve amplitude. This suggests that,
in theory, the most precise systemic velocity may be obtained by averaging
systemic velocities which are estimated from radial velocities measured from
individual Balmer lines. However, in some cases (e.g., for faint stars), it may 
make more sense to measure radial velocities by cross-correlating more than one 
Balmer line with a reference spectrum. Due to the fact that Balmer line velocity
curves have different shapes and amplitudes, we expect that such measurements
may be more uncertain.

To estimate the level of uncertainty introduced by using more than one Balmer
line in a cross-correlation, in Figure~\ref{fig3} we show the standard deviation
of radial velocities of various Balmer lines as a function of phase for RR Lyrae
stars with $V$-band amplitudes ${\rm A_V=0.6,1.0,1.4}$ mag. In general, the
uncertainty in velocity introduced by using more than one Balmer line in a
cross-correlation is lowest for phases earlier than 0.6 ($\lesssim4$ {\kms} for
${\rm A_V}=1$ mag), and it decreases with increasing $V$-band amplitude.

\section{${\rm A_{g}}$ and ${\rm A_{r}}$ vs.~${\rm A_V}$ correlations\label{sdss_correlations}}

Equations~\ref{metallic_lines} to~\ref{Hgamma} describe the amplitude
correlations between radial velocity curves and the $V$-band light curve. To
enable use of these equations with RR Lyrae stars that have been observed in the
Sloan Digital Sky Survey (SDSS; \citealt{yor00}) $g$- and $r$-band filters, we
also derive the correlations between the light curve amplitudes of RR Lyrae
stars observed in SDSS $g$ and $r$ bands and the Johnson $V$ band.

In deriving the correlations, we use the best-fit templates and light curve
parameters of 379 $ab$-type RR Lyrae stars observed in SDSS Stripe 82
(see Table 2 in \citealt{ses10a}). We synthesize each RR Lyrae star's Johnson
$V$-band light curve from the best-fit SDSS $g$- and $r$-band light curves
($g_{LC}(\phi)$ and $r_{LC}(\phi)$) using a photometric transformation given by
Equation 10 from \citet{ive07}:
\begin{eqnarray}
V_{LC}(\phi) = g_{LC}(\phi) + 0.0688gr_{LC}(\phi)^3 -0.2056gr_{LC}(\phi)^2 -0.3838gr_{LC}(\phi) - 0.053,
\end{eqnarray}
where $gr_{LC}(\phi) = g_{LC}(\phi)-r_{LC}(\phi)$, and $\phi\in[0,1\rangle$ is
the phase of pulsation. The $V$-band amplitude is then simply measured from the
synthesized light curve.

Having measured $V$-band amplitudes (${\rm A_V}$), we can now study how they
correlate with SDSS $g$- and $r$-band amplitudes. We find that, within
$\sim0.02$ mag of uncertainty (rms scatter), the $V$-band amplitudes can be
calculated as ${\rm 0.9A_{g}}$ and ${\rm 1.21A_{r}}$, where ${\rm A_{g}}$ and
${\rm A_{r}}$ are SDSS $g$- and $r$-band light curve amplitudes of RR Lyrae
stars. The 1.21 factor is also obtained for the Johnson $R$ band, by comparing
Johnson $V$- and $R$-band light curves of RR Lyrae stars V3, V4, V5, and V10
observed in NGC 5053 by \citet{fer10}.

\section{Systemic velocity and its uncertainty\label{systemic_velocity}}

In Section~\ref{templates}, the templates were defined to have a zero point at
phase of 0.5, since we assumed the radial velocity at phase of 0.5 to be equal
to the systemic velocity. Using this definition, the observed radial velocity
$v_{obs}$ can be calculated as
\begin{equation}
v_{obs}(\Phi_{obs}) = A_{rv}T(\Phi_{obs}) + v_\gamma,
\end{equation}
where ${\rm A_{rv}}$ is the amplitude of the radial velocity curve
(calculated using Equations~\ref{metallic_lines} to~\ref{Hgamma}),
${\rm T(\Phi_{obs})}$ is the template radial velocity curve, ${\rm \Phi_{obs}}$
is the phase of observation, and ${\rm v_{\gamma}}$ is the systemic velocity.

We use the following equation to model the uncertainty in the systemic velocity
(${\rm \sigma_{v_\gamma}}$)
\begin{equation}
\sigma_{v_\gamma} = (A_{rv}+\sigma_{fit})^2[\sigma^2_{template}(\Phi_{obs}) + (0.1k)^2],
\end{equation}
where $k$ is the slope of a template between phases 0.4 and 0.6 (1.16, 1.47,
1.54, and 1.42 for the metallic, ${\rm H\alpha}$, ${\rm H\beta}$, and
${\rm H\gamma}$ line templates, respectively). In the above equation, the
$\sigma^2_{template}(\Phi_{obs})$ term describes the rms scatter in the template
at the phase of observation (bottom plots in Figure~\ref{fig1}), and the
$(0.1k)^2$ term describes the error introduced by the uncertainty in phase at
which the radial velocity is equal to the systemic (center-of-mass) velocity.
These two terms are given in units of velocity amplitude, so they scale with
${\rm A_{rv}}$. To take into account the uncertainty in
Equations~\ref{metallic_lines} to~\ref{Hgamma}, we add the rms scatter of the
${\rm A_{rv}}$ vs.~${\rm A_V}$ fit to our estimate of the velocity amplitude.

The $(0.1k)^2$ term describes our lack of knowledge of the exact phase at which
the radial velocity is equal to the systemic (center-of-mass) velocity. In
Section~\ref{templates}, we assumed this phase to be at 0.5. However,
\citet{oke62} used radial velocity measurements of RR Lyrae star SU Dra and
found the systemic velocity of this star to be consistent with the observed
radial velocity at phase 0.4 (see his Section 6). At this phase, the velocity
curves measured from metallic and ${\rm H\gamma}$ lines intersect (i.e., the
velocity gradient vanishes, see Figure 3 of \citealt{oke62}). Looking at the
radial velocity data for RR Lyrae stars Z Mic and RV Oct shown in Figure 2 of
\citet{pre11}, the metallic and ${\rm H\gamma}$ lines intersect at phases 0.5
and 0.6, respectively. In a different study, \citet{oke66} observed the systemic
velocity of RR Lyrae star X Ari at phase 0.5. Therefore, we conclude that the
uncertainty in the exact phase of systemic velocity is about 0.1. This
uncertainty translates into an uncertainty in systemic velocity, which scales
with the slope of the template between phases 0.4 and 0.6 ($k$), and naturally,
with the amplitude of the velocity curve (${\rm A_{rv}}$). For a typical RR
Lyrae star with a $V$-band amplitude of ${\rm A_V\sim1}$ mag, this uncertainty
introduces an error of about 13 {\kms} into the estimate of systemic velocity 
if the ${\rm H\gamma}$ line is used for radial velocity measurements, or higher
if the ${\rm H\beta}$ or ${\rm H\alpha}$ lines are used. This uncertainty is
only $\sim7$ {\kms} if one uses metallic lines.

\section{Discussion and conclusions\label{discussion}}

\begin{figure}
\epsscale{0.8}
\plotone{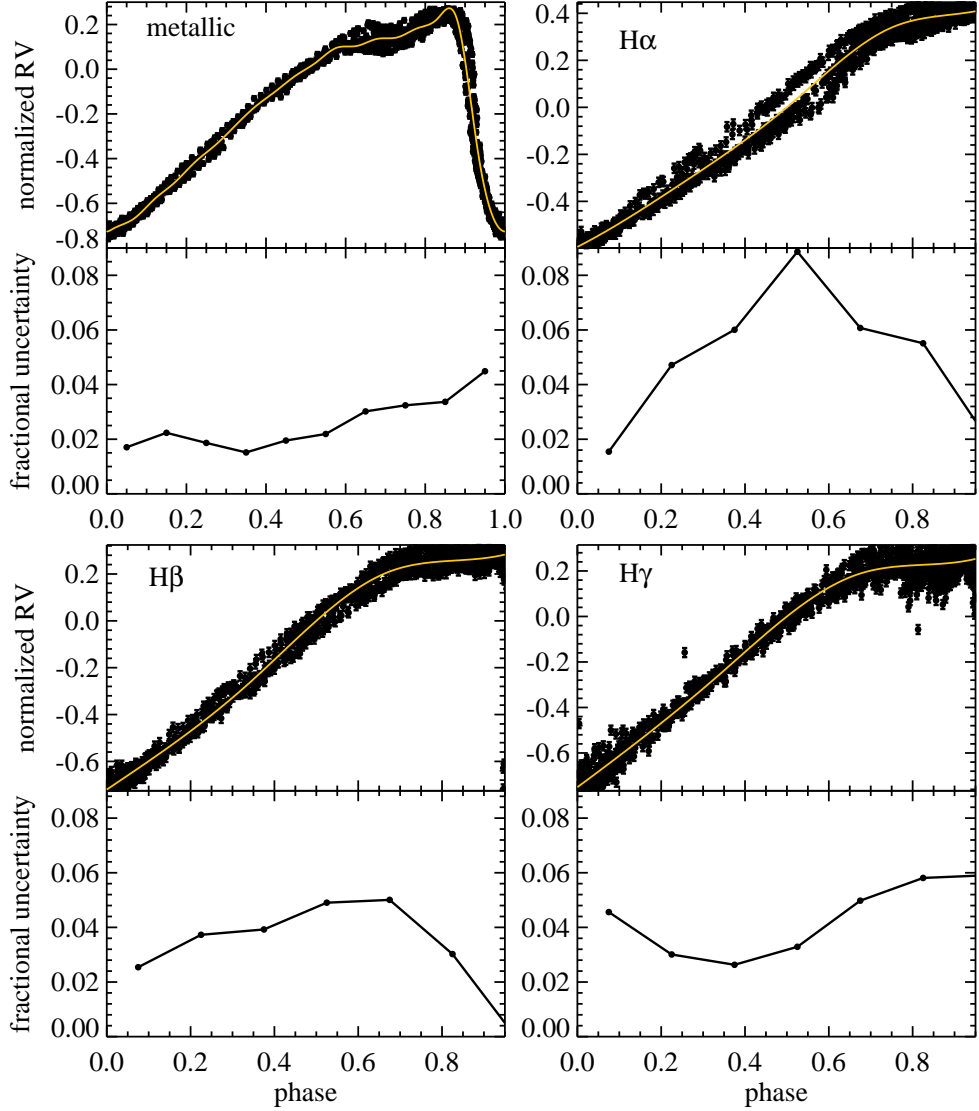}
\caption{
Normalized and phased radial velocity curves obtained from measurements of
metallic lines ({\em top left panel}), ${\rm H\alpha}$ line ({\em top right
panel}), ${\rm H\beta}$ line ({\em bottom left panel}), and ${\rm H\gamma}$ line
({\em bottom right panel}). In each panel, the solid line shows a B-spline curve
interpolated through normalized radial velocities (i.e., the template radial
velocity curve). The template and the normalized radial velocities are offset so
the velocity at phase 0.5 (the phase at which the radial velocity is presumed to
be equal to the systemic velocity, see Section~\ref{systemic_velocity}) is equal
to 0. The connected points in the bottom plot of each panel show the rms scatter
in the template velocity curve (in units of ${\rm A_{rv}}$), and represent the
uncertainty in the shape of the template.
\label{fig1}} 
\end{figure}

\begin{figure}
\epsscale{1.0}
\plotone{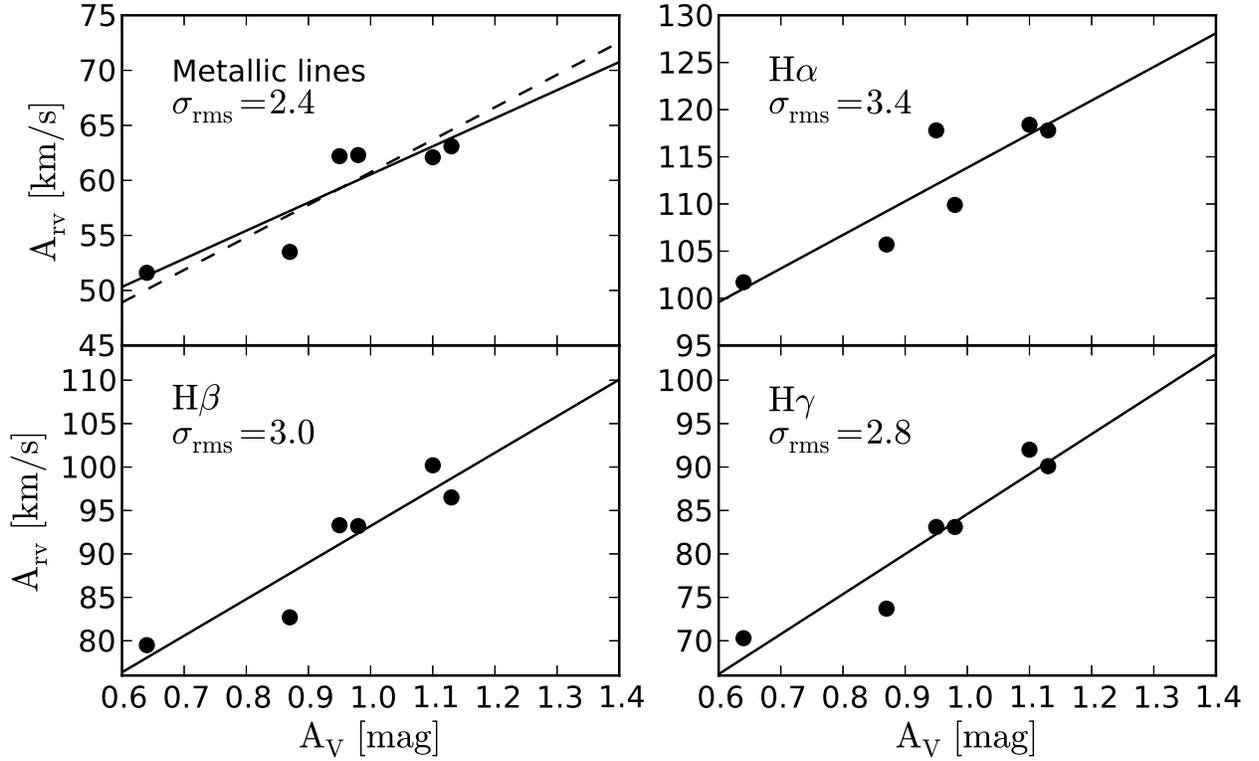}
\caption{
Correlations between radial velocity amplitude, $A_{rv}$, and $V$-band light
curve amplitude, $A_V$, for radial velocities measured from metallic lines
({\em top left panel}), ${\rm H\alpha}$ line ({\em top right panel}),
${\rm H\beta}$ line ({\em bottom left panel}), and ${\rm H\gamma}$ line
({\em bottom right panel}). The solid lines show an unweighted linear
least-squares fit to plotted points (Equations~\ref{metallic_lines}
to~\ref{Hgamma}), and ${\rm \sigma_{rms}}$ indicates the rms scatter of the fit
in {\kms}. The dashed line in the top left panel shows the fit obtained by L91
(see Section~\ref{correlations} for more details). Even though Liu's sample
spans a much greater range of $V$-band amplitudes than ours and has more stars,
the two fits agree within uncertainties (rms scatter of $\sim2.5$ {\kms}),
indicating the our fit's robustness.
\label{fig2}} 
\end{figure}

\begin{figure}
\epsscale{0.5}
\plotone{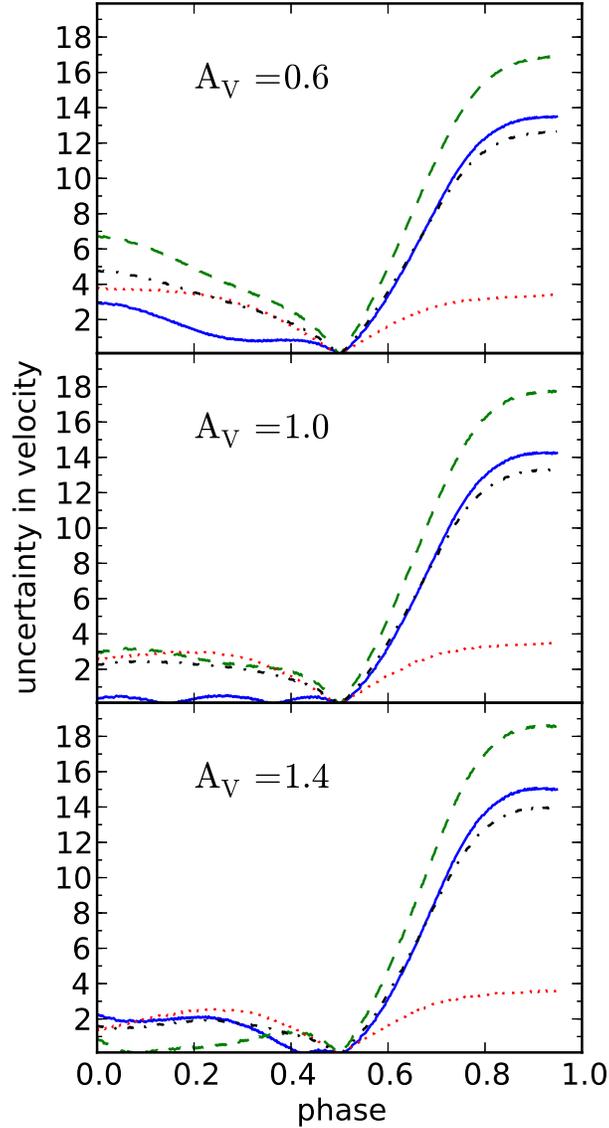}
\caption{
These panels show the uncertainty in velocity (in {\kms}) introduced by using
more than one Balmer line in a cross-correlation, for RR Lyrae stars with
different $V$-band light curve amplitudes (${\rm A_V}$). The solid line shows
the uncertainty when the ${\rm H\alpha}$ and ${\rm H\beta}$ lines are used, the 
dashed line shows the uncertainty when the ${\rm H\alpha}$ and ${\rm H\gamma}$
lines are used, the dot-dashed line shows the uncertainty when ${\rm H\beta}$
and ${\rm H\gamma}$ lines are used, and the dotted line shows the uncertainty
when all three lines are used. In general, the uncertainty in velocity
introduced by using more than one Balmer line in a cross-correlation is lowest
for phases earlier than 0.6 ($\lesssim4$ {\kms} for ${\rm A_V}=1$ mag), and it
decreases with increasing $V$-band amplitude.
\label{fig3}} 
\end{figure}

We have presented template radial velocity curves of $ab$-type RR Lyrae stars
constructed from high-precision measurements of ${\rm H\alpha}$, ${\rm H\beta}$,
and ${\rm H\gamma}$ lines. Compared to the template radial velocity curve (solid
line in the top right plot of Figure~\ref{fig1}), the shape of observed
${\rm H\alpha}$ velocity curves varies the most (up to $\sim10$ {\kms}). The
observed metallic lines curves show the least amount of variability ($\sim3\%$
of ${\rm A_{rv}}$ or $\sim2$ {\kms} for a RR Lyrae star with ${\rm A_V} = 1.0$
mag), followed by ${\rm H\gamma}$ and ${\rm H\beta}$ velocity curves ($\sim3\%$
to $4\%$ of ${\rm A_{rv}}$ or $\sim4$ {\kms}). The fluctuation in the shape of
metallic, ${\rm H\gamma}$ and ${\rm H\beta}$ velocity curves is consistent with
measurement errors, and implies little or no variation in the shape of velocity
curves over a wide range of $V$-band amplitudes (0.6 to 1.1 mag), even for RR
Lyrae stars that are undergoing Blazkho modulations.

We have found tight correlations (rms scatter $<3.4$ {\kms}) between amplitudes
of Balmer lines velocity curves (${\rm A_{rv}}$) and $V$-band light curves
(${\rm A_V}$). A similar correlation, but for metallic lines, was first reported
by L91. We also find that the ratio of amplitudes of Balmer versus metallic
lines velocity curves decreases towards shorter wavelengths, and is 1.90, 1.54, 
and 1.39 for ${\rm H\alpha}$, ${\rm H\beta}$, and ${\rm H\gamma}$ lines. This
pattern basically follows the depth of formation of lines; the ${\rm H\gamma}$
line forms closer to the photosphere than the ${\rm H\alpha}$ line, and
therefore has smaller variations in the radial velocity.

The correlations derived in this paper have the potential to significantly
improve the precision of systemic velocities of RR Lyrae stars estimated from
a small number of radial velocity measurements of Balmer lines. For example,
Equation~\ref{Hgamma} indicates that previous studies that used the
${\rm H\gamma}$ velocity curve of X Ari (${\rm A_{rv}\sim85}$ {\kms},
${\rm A_V\sim1.0}$ mag) may have introduced up to $\sim15$ {\kms} of uncertainty
in their estimates of systemic velocities for RR Lyrae stars with $V$-band
amplitudes of 0.6 or 1.3 mag. Such uncertainties can be now be eliminated by
using Balmer line radial velocity templates and Equations~\ref{Halpha}
to~\ref{Hgamma} presented in this work. 

We find the dominant source of uncertainty in the systemic velocity to be the
phase at which the radial velocity is equal to the systemic velocity. We have
estimated this uncertainty in phase to be about 0.1. For a RR Lyrae star with a
$V$-band amplitude of ${\rm A_V=1.0}$ mag, this uncertainty in phase translates
into a velocity uncertainty of about 13 {\kms}. A repeat of the analysis by
\citet{oke62} (see his his Section 6) on radial velocity data used in our work
may provide more insight into this fundamental uncertainty.

Regrettably, we could not obtain radial velocity measurements of the
${\rm H\delta}$ line (due to low signal-to-noise ratio in echelle data; George
W. Preston, private communication), and were thus unable to construct a template
curve or establish a correlation between the ${\rm H\delta}$ velocity amplitude 
and the $V$-band light curve amplitude. This is quite unfortunate as this line
is often used in radial velocity measurements of RR Lyrae stars and is usually
accessible (along with ${\rm H\beta}$ and ${\rm H\gamma}$ lines) to most
low-resolution spectrographs, such as DBSP \citep{og82} or LRIS \citep{oke95}.
A spectroscopic survey that would provide velocity measurements of the
${\rm H\delta}$ line for several RR Lyrae stars would be very useful, and would
allow the extension of this work to the ${\rm H\delta}$ line.

By deriving correlations between light curve amplitudes of RR Lyrae stars
observed in SDSS $g$ and $r$ bands and the Johnson $V$ band
(Section~\ref{sdss_correlations}), we have expanded the applicability of
Equations~\ref{Halpha} to~\ref{Hgamma} to a large number of already or soon to
be observed RR Lyrae stars. For example, there are $\sim380$ $ab$-type RR Lyrae
stars with measured light curve parameters (period, epoch of maximum light,
amplitudes in SDSS $ugriz$ band, etc.)~that have been observed in SDSS stripe 82
\citep{ses10a}, a large fraction of which have spectroscopic observations
\citep{del08}. While the systemic velocities of these stars have already been
measured \citep{del08}, these measurements could be improved using
Equations~\ref{Halpha} to~\ref{Hgamma}, leading to better estimates of kinematic
properties of the Galactic halo.

Then there are multi-epoch, wide-area surveys such as the Lincoln Near-Earth
Asteroid Research (LINEAR; \citealt{ses11}) and Palomar Transient Factory (PTF;
\citealt{law09}), which are currently finding thousands of RR Lyrae stars within
30 kpc from the Sun \citep{sesar11}. Some of these stars have previously been
spectroscopically observed by SDSS, and many more are expected to be observed by
the upcoming low-resolution spectroscopic survey LEGUE ($R=1800$;
\citealt{den12}). By using light curve parameters of RR Lyrae stars from LINEAR
and PTF, radial velocities from LEGUE, and Equations~\ref{Halpha}
to~\ref{Hgamma}, one could easily construct a sample of RR Lyrae stars with
precise positions and 3D kinematics, and use them to search for halo
substructures within 30 kpc from the Sun.

And finally, the Large Synoptic Survey Telescope (LSST; \citealt{ive08}) is
expected to be able to observe RR Lyrae stars as far as 360 kpc from the Sun
\citep{olu12}. With RR Lyrae stars detected at these distances, LSST will be
able to search for halo streams and dwarf satellite galaxies within a
significant fraction of the Local Group. Due to faintness of these distant RR
Lyrae stars ($V\lesssim23.4$), the spectroscopic followup will most likely
involve observations of Balmer lines at low resolutions ($R\sim1500$), making
the tools presented in this work a natural choice for the measurement of their
systemic velocities.

\acknowledgments

B.S. would like to acknowledge Judith G. Cohen, Shrinivas R. Kulkarni, and
Carl J. Grillmair for their support, which was made possible by NSF grants
AST-0908139 (to J.G.C.) and AST-1009987 (to S.R.K.), and a NASA grant
(to C.J.G.). B.S. would also like to thank George W. Preston for kindly
providing his data that made this study possible.

\begin{deluxetable}{ccccccccccc}
\tabletypesize{\scriptsize}
\tablecolumns{7}
\tablewidth{0pc}
\tablecaption{ASAS RR Lyrae Stars\label{ASAS_RR}}
\tablehead{
\colhead{Star} & \colhead{R.A.(J2000)} & \colhead{Dec(J2000)} &
\colhead{${\rm A_V}^a$} & \colhead{${\rm V_{max}}^b$} & \colhead{Blazhko$^c$} &
\colhead{$N_{obs}$} &
\colhead{Metallic ${\rm A_{rv}}^d$} &
\colhead{${\rm H\alpha}$ ${\rm A_{rv}}^d$} &
\colhead{${\rm H\beta}$ ${\rm A_{rv}}^d$} &
\colhead{${\rm H\gamma}$ ${\rm A_{rv}}^d$} \\
\colhead{} & \colhead{(hr $\arcmin$ $\arcsec$)} &
\colhead{($\arcdeg$ $\arcmin$ $\arcsec$)} & \colhead{(mag)} & \colhead{(mag)} &
\colhead{} & \colhead{} & \colhead{({\kms})} & \colhead{({\kms})} & \colhead{({\kms})} &
\colhead{({\kms})}
}
\startdata
Z Mic  & 21 16 22.71 & -30 17 03.1 & 0.64 & 11.32 & Yes & 166 & 51.6 & 101.7 & 79.5  & 70.3 \\
CD Vel & 09 44 38.24 & -45 52 37.2 & 0.87 & 11.66 & Yes & 193 & 53.5 & 105.7 & 82.7  & 73.7 \\
WY Ant & 10 16 04.95 & -29 43 42.4 & 0.95 & 10.37 & No  & 130 & 62.2 & 117.8 & 93.3  & 83.1 \\
DT Hya & 11 54 00.18 & -31 15 40.0 & 0.98 & 12.53 & No  &  96 & 62.3 & 109.9 & 93.2  & 83.1 \\
XZ Aps & 14 52 05.43 & -79 40 46.6 & 1.10 & 11.94 & No  & 241 & 62.1 & 118.4 & 100.2 & 92.0 \\
RV Oct & 13 46 31.75 & -84 24 06.4 & 1.13 & 10.53 & Yes & 193 & 63.1 & 117.8 & 96.5  & 90.1
\enddata
\tablenotetext{a}{$V$-band light curve amplitude from ASAS.}
\tablenotetext{b}{Maximum light in the $V$-band from ASAS.}
\tablenotetext{c}{Blazhko RR Lyrae stars according to \citet{sf07}.}
\tablenotetext{d}{Best-fit radial velocity amplitude.}
\end{deluxetable}

\clearpage

\clearpage

\clearpage

\end{document}